\documentstyle[11pt]{article}
\begin{document}
\newcommand{\BE}{\begin{equation}}
\newcommand{\EE}{\end{equation}}
\newcommand{\BA}{\begin{eqnarray}}
\newcommand{\EA}{\end{eqnarray}}
\newcommand{\de}{{\rm d}}
\vspace*{25mm}
\begin{center}
{\LARGE\bf Autonomous Renormalization of $\bf\Phi^4$\vspace{5mm}\\
in Finite Geometry}
\vspace{30mm}\\
{\Large U. Ritschel}
\vspace{10mm}\\
{\large\it
Fachbereich Physik, Universit\"at GH Essen, D-45117 Essen
(F R Germany)}
\vspace{25mm}\\
\end{center}
{\bf Abstract:} The autonomous renormalization of the
$O(N)$-symmetric
scalar theory is based on an infinite re-scaling of constant
fields, whereas finite-momentum modes remain finite. The natural
framework
for a detailed analysis of this method is a system of finite size,
where
all non-constant modes can be integrated out perturbatively and
the
constant mode is treated by a
saddle-point approximation in the thermodynamic limit. Our
calculation provides a better understanding of the properties of
the effective
action and corroborates earlier findings concerning a heavy Higgs
particle
at about 2 TeV \cite{cons,rodr}.
\newpage
Seven years ago Stevenson and Tarrach discovered the autonomous
renormalization \cite{auto,onsy} in the framework of the
Gaussian variational
approximation \cite{stev0} for the
$\phi^4$-theory in $3+1$ dimensions. More recently,
Consoli et al. \cite{cons} and
Iba\~nez-Meier et al. \cite{rodr} where able to derive predictions
for
the Higgs mass from the autonomously renormalized
$\phi^4$-model under the natural assumption that
the Higgs sector is massless in the symmetric phase.

The main ingredient in the autonomous
renormalization (AR) are an infinite
re-scaling of the
constant mode of the field and an UV-flow of the bare coupling
constant that cancels
the leading logarithmic divergences.
When applied to the effective
potential in Gaussian \cite{auto}, one-loop \cite{cons,rodr}, or
more sophisticated variational approximations \cite{stan},
the procedure gives always a finite answer.
Problems occur if one
attempts to calculate the full effective action with the AR.
It turns out
that infinities in the kinetic terms of the action are not removed
by the AR \cite{dive}.

Recently, Consoli and Stevenson \cite{cost} demonstrated
that the solution to this
apparent dilemma is a wavefunction renormalization that
distinguishes
between constant and non-constant
or finite-momentum (FM) modes. Constant fields have
to be
re-scaled by an infinite factor, whereas FM fields remain
finite. In this respect the procedure
is quite different from conventional
renormalization, where all modes are
renormalized by a common Z-factor,
but it does not violate any fundamental principles otherwise
\cite{cost}.
Further, it turns out that the
interaction between FM modes is suppressed by powers
of $1/\log \Lambda$
and all the non-trivial structure, especially the symmetry
breaking
form of the effective potential, is caused by the self-interaction
of the constant modes and their coupling to the FM
modes.

The aim of the present paper is a closer analysis of the AR of the
effective action.
The most natural environment to investigate a renormalization that
distinguishes between modes
is a system of finite size, where one deals with
a discrete set of functions.
It is well known from the theory of second-order
phase transitions that in a finite geometry
the constant mode has to be treated non-perturbatively in order
to avoid spurious IR-divergences, while the FM fields may be
integrated out perturbatively \cite{ziju}.
Thus, one
may largely rely on techniques developed in statistical
field theory, once one has derived
an effective theory for the (renormalized) constant mode with the
help
of the AR.

In the following we are concentrating on a scalar field
in a four dimensional euclidean space, with
volume $\Omega=L^3/T$, which describes a system
in a finite spatial volume $L^3$ at temperature $T$.
To begin with, we consider the case $N=1$.
Concerning the perturbative evaluation of the FM
modes the generalization to $N\neq 1$ is
straightforward. Only the constant angular variables (Goldstone
modes)
need a separate treatment.
This will be discussed in more detail below.

In terms of Fourier amplitudes the field is given by
\BE
\phi(x) = \frac{1}{\sqrt{\Omega}}
\sum_{k} \phi_k \>{\rm e}^{{\rm i}\,p\cdot x}\>,
\EE
where $p_0=2\pi k_0\,T$, $p_i=2\pi k_i/L$, for
$i=1,2,3$ and $k_0,k_i \in Z$.
For the zero mode we choose the notation
$\phi_{k=0}\equiv \sqrt{\Omega}\phi_0$, such that
$\phi_0=1/\Omega\int\de^4x \,\phi(x)$.
Then the generating functional reads (up to unimportant
normalization
factors)
\BE\label{genfu}
Z(j) \sim \int {\rm d}\phi_0 \>\Pi' {\rm d}\phi_k
\>{\rm exp}\left(-S(\phi_0,\phi_k)-\Omega\,
j_0 \phi_0-\sum'j_k\phi_{-k}\right)
\EE
where the prime denotes products and sums over all
$k\in Z^d\setminus \{0\}$.
The action in (\ref{genfu}) is given by
\BE\label{action}
S(\phi_0,\phi_k)= \Omega\> U_{{\rm cl}}(\phi_0)
+\frac12\,\sum ' \left[p^2+\frac{g}{2}
\phi_0^2\right]\,\phi_k\phi_{-k}\>+\>{\cal O}(\phi_k^3)\>.
\EE
with
\BE
p^2=4\pi^2\left(k_0^2\,T^2+\frac{\vec k^2}{L^2}\right)
\EE
and with the classical potential
\BE
U_{\rm cl}(\phi_0)=\frac12\> t  \> \phi_0^2+\frac{g}{4!}\>
\phi_0^4\>.
\EE

The prime modes can be integrated out in a one-loop-type
procedure,
i.e., taking into account only quadratic terms in $\phi_k$.
The result of the integration reads
\BE\label{zofj}
Z[j]\sim \int_{-\infty}^{\infty}{\rm d}\phi_0\>
{\rm exp}\left[-
\Omega\left(s(\phi_0)+{\cal J}(\phi_0)-j_0\phi_0\right)+
\frac12\sum'j_k\,G_k j_{-k}\right]
\EE
where
\BE\label{eione}
{\cal J}(\phi_0)=\frac{1}{2\Omega}\sum'\log\left[1+
\frac{g\,\phi_0^2}{2\,p^2}\right]
\EE
and the propagator is given by
\BE\label{prop}
G_k=\frac{1}{p^2+g\phi^2_0/2}\>.
\EE
Concerning the UV-behavior, the term ${\cal J}(\phi_0)$ is
equivalent
to what is usually called
$I_1$ in the infinite volume procedure \cite{stev0}; the finite
extent of the system does not change the behavior at distances
$d\ll L$.

In order to extract the UV-divergences from (\ref{eione}),
it is sufficient to analyze the derivative
of ${\cal J}$ - the $\phi_0$-independent infinities are not
relevant
for the generating functional -,  which is given by
\BE\label{inot}
{\cal I}\equiv\frac{{\rm d}{\cal  J}}{{\rm d}\phi_0}
=\frac{g \phi_0}{2\Omega}\sum' \frac{1}{p^2+g\,\phi_0^2/2}
= \frac{g\phi_0}{2\Omega}
\int_0^{\infty}\de s\> \left[A^3\left(
\frac{4\pi^2 s}{L^2}\right)\,A\left(4\pi^2 s\,T^2\right)-1\right]
{\rm e}^{-s g \,\phi_0^2/2}
\EE
with
\BE
A(x)=\sum_{n=-\infty}^{\infty}{\rm e}^{-n^2 x}\>.
\EE
In the integral on the right hand side of (\ref{inot}),
the UV-divergence of the original sum over momenta is reflected by
a
pole of the integrand at small s, where the function $A(x)$
behaves
as $(\pi/x)^{\frac12}$. By adding and subtracting terms, ${\cal
I}$ can be
written as
\BE
{\cal I }= \frac{g \phi_0}{2} \left( K_{\infty}+K_{F}\right)\>,
\EE
where
\BE
K_{\infty}= \frac{1}{16\pi^2}\int_0^{\infty}\>\frac{\de s}{s^2}
\>{\rm e}^{-s g\, \phi_0^2/2}
\EE
contains the UV-divergence and
\BE
K_{F}= \frac{1}{\Omega}
\int_0^{\infty}{\rm d}s\>
\left[A^3\left(\frac{4\pi^2 s}{L^2}\right)\,A\left(4\pi^2
s\,T^2\right)
-1-\frac{\Omega}{16\pi^2 s^2}\right]
\>{\rm e}^{-s g\, \phi_0^2/2}
\EE
is finite. To determine the UV-flows of the bare parameters,
it is favorable to
regularize $K_{\infty}$ by a cutoff $\Lambda^{-2}$ at the lower
bound
of the integral.
Carrying out the integration gives
\BE
K_{\infty}=
\frac{1}{16\pi^2}  \left\{ \Lambda^2
+\frac{g\phi_0^2}{2}\left[C-1+
\log\left(\frac{g\phi_0^2}{2\Lambda^2}\right)
\right]+{\cal O}\left(\phi_0^4/\Lambda^2\right)\right\}\>,
\EE
where $C$ stands for Euler's constant. Eventually, by
integrating once with respect to $\phi_0$, one obtains
\BE\label{jay}
{\cal J}=
\frac{1}{64\pi^2} \left\{ g\>\phi_0^2\Lambda^2
+\frac{g^2\phi_0^4}{4}\left[C-\frac32+
\log\left(\frac{g\phi_0^2}{2\Lambda^2}\right)
\right]\right\}  +\frac{g}{2} \int^{\phi_0}{\rm d}\phi\>\phi\>
K_F(\phi)+
{\cal O}\left(\phi_0^6/\Lambda^2\right).
\EE
The most systematic way to obtain a finite generating functional
is a renormalization-group formulation
as first employed in this context by Consoli et al. \cite{asym}.
In order
to determine the UV-flows of the parameters of
the system, it is
sufficient to solve
\BE\label{rgeqn}
\left(
\Lambda\> \frac{\partial}{\partial\Lambda} + \beta\>
\frac{\partial}
{\partial g}-\frac{\eta}{2}\>  \phi_0\> \frac{\partial}{\partial
\phi_0}
-\eta_2 \> t\>  \frac{\partial}{\partial t}\right)
\left[s(\phi_0;g,t)+{\cal J}(\phi_0;g)\right] =0
\EE
with Wilson functions
\BE\label{beta}
\beta = \Lambda\frac{\de\,g}{d\,\Lambda}\>,\quad\eta
=-2\Lambda\> \frac{\de
\log\phi_0}{\de \Lambda}\>,\quad{\rm and}\quad
\eta_2=-\Lambda\frac{\de \log t}{\de\Lambda}  \>.
\EE
One solution to (\ref{rgeqn}) - the other is the conventional
perturbative one \cite{cons} - is determined by the
relations
\BE\label{wils}
\eta=\frac{\beta}{g}\>,\quad t(\eta+\eta_2)=
\frac{g\Lambda^2}{16\pi^2}\>,\quad {\rm and} \quad
\beta=-\frac{3\,g^2}{16\pi^2}\>,
\EE
which, in turn, yield the explicit solutions
\BA\label{resu}
g(\Lambda)&=&\frac{16\pi^2}{3\log(\Lambda/K)}\>,\nonumber \\
\phi_0^2(\Lambda)&=&z_0\,\Phi^2\log(\Lambda/K)\>,\nonumber \\
\quad t(\Lambda)&=&-\frac{\Lambda^2}{6\log(\Lambda/K)}
+\frac{c}{\log(\Lambda/K)}\>.
\EA
In the above formulae, $\Phi$ is the renormalized expectation
value of the
field. The parameters $K$, $z_0$, and $c$
have to be determined by normalization conditions.
For instance one finds $c=0$ when the system has
massless excitations in the symmetric state.
It is this case, which will be considered in the following.

Inserting (\ref{resu}) in the exponent of (\ref{zofj}) we find for
$\sigma(\Phi)\equiv U_{\rm cl}(\phi_0) + {\cal J}(\phi_0)$,
\BA\label{sigma}
\sigma(\Phi) & = &\frac{\pi^2}{9}z_0^2
\Phi^4 \log\left(\Phi^2/\mu^2\right)\nonumber\\
&
+&\frac{1}{2\,\Omega}\; \int_0^{\infty}\frac{\de s}{s}\left[
\,A^3\left(
\frac{4\pi^2s}{L^2}\right)\,A\left(4\pi^2s\,T^2\right)
-1-\frac{\Omega}{16\pi^2s^2}\right]\>
\left(1-{\rm e}^{-8\pi^2
s \,z_0\Phi^2/3}\right)\>,
\EA
where $\mu$ is a new mass parameter, essentially $K$ multiplied by
numerical constants.
The function $\sigma(\Phi)$
may be regarded as a precursor of the effective potential and,
indeed, in certain cases $\sigma(\Phi)$ turns out to be
identical with the effective
potential. In general, however, $\Phi$ is an integration variable
and not the expectation value of the field.
In order to obtain the generating
functional, we have to calculate the integral
\BE\label{zopj}
Z[j]\sim \int_{-\infty}^{\infty} \de\Phi\>{\rm
exp}\left(-\Omega\left(
\sigma(\Phi)-{\rm j}\Phi\right)+\frac12 \sum'j_k G_k
j_{-k}\right)\>,
\EE
where the zero mode of the source term has been appropriately
re-scaled,
and the propagator $G_k$ is given by (\ref{prop}) with
renormalized
squared mass
\BE
\lim_{\Lambda\rightarrow\infty} \frac{g}{2}\phi_0^2
= \frac{8\pi^2}{3}z_0\Phi^2\> .
\EE

As we are actually interested in the limit of large space volume
of (\ref{zopj}), we may consider $\sigma(\Phi)$
for $L\rightarrow\infty$, where it takes the form
\BE\label{effpo}
\sigma(\Phi) = \frac{\pi^2}{9}
z_0^2 \Phi^4 \log\left( \Phi^2/\mu^2\right)
+\frac{1}{32\pi^2} \int_0^{\infty} \frac{\de s}{s^3} \left[
A\left(\frac{1}{4\,s\,T^2}\right)-1\right]\>\left(1-{\rm
e}^{-8\pi^2
s\,z_0\,\Phi^2/3}\right)+{\cal O}(1/\Omega)\, .
\EE
Neglecting the terms that vanish relatively for large volume,
$\sigma(\Phi)$ is identical with the finite-temperature
effective potential
as obtained when the AR is applied in infinite volume from the
start
\cite{hajj}. This means that $\sigma(\Phi)$ has a form typical
for a first-order phase transition. It shows a single global
minimum if
the temperature is higher than some critical value $T_c$. For $T <
T_c$, one finds two degenerate minima located symmetrically with
respect
to the origin. At the critical value there are three degenerate
minima.

Further, as in Ref. \cite{ziju},
we use the saddle-point approximation for the $\Phi$-integration
in (\ref{zopj}).
The situation is particularly simple if the potential
$\sigma(\Phi)$
is convex everywhere, which is the case when $T$ is well above
$T_c$.
Only then there exists a unique saddle
point for each value of ${\rm j}$ and
the generating functionals can be calculated
explicitly. For instance for the effective action we find
\BE\label{effa}
\Gamma[\bar\phi] = \frac12
\sum' \left( p^2 + \frac{8\pi^2}{3}z_0
\bar\Phi^2\right) \bar\phi_k\bar\phi_{-k} + \Omega
\,\sigma(\bar\Phi)\>,
\EE
and the effective potential is given by
\BE\label{efpo}
U_{{\rm eff}}(\bar\Phi)=\frac{1}{\Omega}\>
\Gamma[\bar\phi=const.]=\sigma(\bar\Phi)\>,
\EE
where $\bar\phi$ is the expectation value of the field, and
$\bar\phi_k$
and $\bar\Phi$ are the amplitudes of FM modes and renormalized
constant modes, respectively.
(\ref{effa}) is consistent with
the result of \cite{cost}, and the
simple relation (\ref{efpo}) holds wherever $\sigma(\Phi)$ is
convex
\cite{jezi}.

The situation becomes
more complicated, however,
when $\sigma(\Phi)$ has concave portions or even
degenerate minima,
leading to the well-known non-analytical
behavior of effective action and effective
potential in such circumstances.
In case $\sigma(\Phi)$ has two minima at $\pm \Phi_m$ and one
minimum at the origin with $\sigma(\Phi_m)\le\sigma(0)=0$,
the ``free energy" for
small constant ${\rm j}$ and large volume is given by
\BE\label{free}
W({\rm j})=\log\left[Z({\rm j})/Z(0)\right]=
\log\left[\frac{\cosh(\Omega\,{\rm j}\,\Phi_m)+\xi}{1+\xi}\right]
\EE
with
\BE
\xi= \frac12
\left(\frac{\sigma''(\Phi_m)}{\sigma''(0)}\right)^{\frac12}
{\rm e}^{\Omega\,\sigma(\Phi_m)}\>.
\EE
For the expectation of $\Phi$ we find
\BE\label{deri}
\langle\Phi\rangle=
\frac{1}{\Omega}\left.
\frac{\de W}{\de {\rm j}}\right|_{{\rm j}\rightarrow 0^{\pm}}=
\frac{\Phi_m\sinh(\Omega
{\rm j}\Phi_m)}{\cosh(\Omega {\rm j} \Phi_m)+\xi}\>.
\EE
This means that for any large but finite $\Omega$ the expectation
value vanishes for ${\rm j}\rightarrow 0^{\pm}$.
However, in the limit $\Omega \rightarrow \infty $ {\it and} ${\rm
j}\rightarrow 0^{\pm}$ we either obtain
$\langle \Phi\rangle=0$ for $\Omega \;{\rm j}\rightarrow 0$
(corresponding to a supercooled state) or
$\langle \Phi\rangle=\pm\Phi_m$
for  $\Omega\; {\rm j}\rightarrow \pm\infty$, which is the
symmetry-breaking
solution we are actually interested in.

A reasonable definition of the mass of the single-particle
excitation is provided by the second derivative of
$\sigma(\bar\Phi)$ in the
limit $\bar\Phi\rightarrow \Phi_m$ from
above. For the moment, we are interested in the mass for
temperature $T=0$. So we may neglect the contribution
from the integral in (\ref{effpo}).
Implementing the normalization condition
\BE\label{norm}
\sigma''(\Phi_m)= \left.(G_{k=0})^{   {-}1}\right|_{\Phi=\Phi_m}
\EE
we find $z_0=3$, which, in turn, yields
\BE\label{mhig}
M^2=8\pi^2\Phi_m^2
\EE
for the mass of single-particle excitation, the Higgs mass.

Going from $N=1$ to the general case $N\neq 1$,
the FM modes are integrated as before.
The UV-flows of the parameters acquire some $N$-dependence
compared
to (\ref{resu}), for instance
\BE
g(\Lambda)=\frac{48\pi^2z_0}{(N+8)\log(\Lambda/K)}\>.
\EE
The generating functional then takes the form
\BE\label{zopn}
Z[j]\sim \int \de^N\Phi\>{\rm exp}\left(-\Omega\left(
\sigma(\Phi^{\alpha})-
{\rm j}^{\alpha}
\Phi^{\alpha}\right)
+\frac12 \sum'j^{\alpha}_k G^{\alpha\beta}_k
j^{\beta}_{-k}\right)\>,
\EE
where the propagator is given by
\BE\label{propn}
G_k^{\alpha\beta}=\frac{1}{p^2+z_N\Phi^2}\left(\delta^{\alpha\beta
}-
\frac{2z_N\Phi^{\alpha}\Phi^{\beta}}{p^2+3z_N\Phi^2}\right)
\EE
with $\Phi^2=\Phi^{\alpha}\Phi^{\alpha}$ and
$z_N=8\pi^2z_0/(N+8)$,
and the analogue to (\ref{effpo}) takes the
$O(N)$-symmetric form
\BE
\sigma(\Phi^{\alpha})
= \frac{\pi^2z_0^2}{N+8}
\left(\Phi^2\right)^2\>
\log(\Phi^2/\mu^2)+\frac{1}{32\pi^2}
\int_0^{\infty} \frac{\de s}{s^3} \left[
A\left(\frac{1}{4\,s\,T^2}\right)-1\right]\>
\left(1-{\rm e}^{-3z_N\,s\,\Phi^2}\right)\>.
\EE

The reason for new effects compared to $N=1$ are
the angular variables.
If one is interested
in the physics of the Goldstone
bosons, one has to express the $N$ euclidean components of the
field by spherical coordinates and expand
loop integrals in terms
of $p^2/M^2$, i.e., for momenta small compared to the
mass of the radial mode \cite{diehl}.
The result is the nonlinear $\sigma$-model,
which describes the physics of the Goldstone bosons
in the low momentum regime.
Here, on the other hand,
we are interested mainly in the radial excitation.
In order to derive (\ref{zopn}),
we have already integrated out perturbatively
all modes with $p\neq 0$. When the constant modes,
$\Phi^{\alpha}$, are expressed
in terms of $N$-dimensional spherical coordinates, $\Phi^{\alpha}=
\rho\,\Phi^{\alpha}$, the angular integrals in (\ref{zopn})
can be carried out
generating an effective theory for the
constant radial field.
On account of the angular dependence
of the sources, however, this cannot be done
out in full generality.

In order to obtain the generating functional for the special
case of constant
source and expectation value of $\rho$, the FM sources
$j^{\alpha}_k$
are set to zero. Without loss of generality, we let $ {\rm
j}^{\alpha}$
point in the $\Phi^1$-direction: ${\rm j}^{\alpha}=({\rm
j},0,\ldots,
0)$.
Carrying out the angular integration leads to
\BE\label{zopo}
Z({\rm j})\sim \int_{0}^{\infty} \de \rho \,\rho^{N-1}\,
\left(\Omega\,
\rho\,{\rm j}\right)^{1-N/2}\,I_{N/2-1}(\Omega\,\rho\, {\rm j})\>
{\rm e}^{-\Omega\,\sigma(\rho)}
\>,
\EE
where $I_k(x)$ denotes a Bessel function of an imaginary argument.
After the saddle-point approximation for the radial
integration (for $T<T_c$ and small ${\rm j}$)
we obtain as the analogue to (\ref{free}):
\BE
W({\rm j}) =
\log\left[\chi^{1-N/2}\,I_{N/2-1}(\chi)\right]\quad{\rm with}
\quad \chi=\Omega\, {\rm j}\,\rho_m\>,
\EE
where $\rho_m$ denotes the
minimum of the potential $\sigma(\rho)$.
The expectation value of
$\rho$ is given by
\BE\label{deri2}
\langle\rho\rangle=
\frac{1}{\Omega}\left.\frac{\de W}{\de {\rm
j}}\right|_{j\rightarrow 0}=
\rho_m
\left(\frac{I_{N/2-2}(\chi)+I_{N/2}(\chi)}{2\,I_{N/2-1}(\chi)}-
 \frac{N-2}{2\,\chi}\right)\>.
\EE
The function
that multiplies $\rho_m$ in the above formula behaves
qualitatively like
the one occuring in (\ref{deri}), i.e., it tends to zero for
$\chi\rightarrow 0$ and it approaches one for
$\chi\rightarrow\infty$.
Hence, the discussion of the limits of $\Omega\,{\rm j}$ carries
over
from (\ref{deri}) with $\Phi_m$ replaced by $\rho_m$, and this
is fully
consistent with the intuitive picture that the symmetry breaking
is caused by some infinitesimal external ``magnetic field", which
determines the direction of the expectation value
of $\Phi^{\alpha}$.

The next step is to calculate the propagator of the FM modes
from (\ref{zopn}). For this purpose we may
expand to second order in $j^{\alpha}_k$.
The angular integration yields
\BE\label{propr}
G_k^{11}=\frac{1}{p^2+z_N\rho_m^2}
\left(1-f_R(\chi)\frac{2\,z_N\,\rho_m^2}{p^2+3\,z_N\,\rho_m^2}
\right)
\EE
for the radial propagator, where, as above, the radial direction
is the one selected by ${\rm j}^{\alpha}$, and
\BE\label{propt}
G_k^{ij}=\frac{\delta^{ij}}{p^2+z_N\rho_m^2}
\left(1-f_T(\chi)\frac{2\,z_N\,\rho_m^2}{p^2+3\,z_N\,\rho_m^2}
\right)
\qquad
i,\,j=2,\ldots,N
\EE
for the transversal propagator with the functions
\BA
f_R(\chi)&=&\left(\chi^{1-N/2}I_{N/2-1}(\chi)\right)^{-1}
\left.\frac{\de^2}{\de x^2}\left(x^{1-N/2}I_{N/2-1}(x)\right)
\right|_{x=\chi}\nonumber\\
f_T(\chi)&=&\frac{N\,I_{N/2}(\chi)}{(N-1)\,\chi\,I_{N/2-1}(\chi)}
\>.
\EA
In the symmetry-breaking limit $\chi\rightarrow\infty$, one finds
$f_R\rightarrow 1$ and $f_T\rightarrow 0$ leading to free
propagators with
mass $M_R^2=3z_N\rho_m^2$ for the radial field and
$M_T^2=z_N\rho_m^2$
for the $N-1$ transversal fields, respectively.
By imposing the normalization condition analogously to
(\ref{norm})
on the radial propagator, we find again $z_0=3$.
The form of the propagators (\ref{propr}) and (\ref{propt}), as
well as the value of $z_0$ are consistent with Ref.
\cite{cost}.
The result for the Higgs mass is $M_R^2=72\pi^2\rho_m^2/(N+8)$,
which
gives $M_R=1.89$ TeV  for $N=4$ and $\rho_m=0.246$ TeV.

In conclusion, we have analyzed the AR scheme
in a finite geometry.
While in \cite{cost} the zero mode has been regarded
as identical with its expectation value, the major contribution of
our
work is a fully quantum-mechanical treatment of all fields. The
FM modes are integrated out perturbatively, and the constant mode
is treated non-perturbatively.
For $N=1$ our approach provides a complete description of
the effective action, including the non-analytic behavior
of the effective potential.
For general $N$, the transversal excitations (the would-be
Goldstone bosons) remain massive in our calculation. Thus, the
reason for this failure has to be sought in
the treatment of the FM modes.\vspace{5mm}\\
{\bf Acknowledgement:} I should like to thank P. M. Stevenson for
helpful discussions and careful reading of the manuscript.

\end{document}